\documentclass[twocolumn,showpacs,amssymb,aps]{revtex4}
\usepackage{amsmath}
\usepackage{amsfonts}
\usepackage{amssymb}
\usepackage{graphicx}
\usepackage{epsfig}
\usepackage{dcolumn}
\usepackage{bm}

\def\lessim{\lower.5ex\hbox{$\; \buildrel < \over \sim \;$}}
\begin{document} \hbadness=10000
\topmargin -1.4 cm
\oddsidemargin = -1.1cm\evensidemargin = -1.1cm
\preprint{}

\title{Electron-Positron Plasma Drop  Formed by Ultra-Intense Laser Pulses}
\author{Inga Kuznetsova and Johann Rafelski}
\affiliation{Department of Physics, The University of Arizona, Tucson, Arizona, 85721, USA}
\date{September 14, 2011}

\begin{abstract}
We study the initial properties and positron annihilation within a small electron-positron plasma drop  formed by intense laser pulse. Such  QED cascade initiated plasma is, in general, far below the chemical (particle  yield) equilibrium. We find that the available electrons and positrons equilibrate  kinetically,  yet despite relatively high particle density, the electron-positron annihilation is very slow, suggesting a rather long lifespan of the plasma drop.
\end{abstract}


\maketitle

\section{Introduction}

Conversion of the high intensity laser pulse energy into a dense gas of $e^+,e^-$ electron-positron pairs is a topic of current theoretical and, soon, experimental  interest. A QED cascade mechanism producing a rapid conversion of laser pulse energy into pairs was demonstrated in~\cite{Nerush:2010fe} for pulse intensity on the order of $10^{24}$ W/cm$^2$. Considering the known reaction cross sections~\cite{Kuznetsova:2009bq}, subsequent to the electromagnetic cascade process discussed in~Ref.\,~\cite{Nerush:2010fe},  photons escape the small plasma drop, while as we show here, the electromagnetic scattering thermalizes the momentum distribution of this relatively dense electron-positron phase. We thus find a drop of "`thermal"' momentum  equilibrated, but  "`chemical"' yield nonequilibrated electron-positron  plasma with a size as small as a few $\mu$m and an energy content up to a kJ. Such plasma will expand, and lose energy by positron annihilation. We obtain here the rates of energy and particle loss by annihilation.
 
The corresponding initial local energy density is provided by the laser field.  We assume the formation of the plasma drop at rest in the lab frame e.g. invoking symmetric laser pulse collisions triggering QED  cascades. The experimental pulse intensity parameter, defining plasma drop properties, is~\cite{TajMou}
\begin{equation}
a_0 = \frac{eE_0\lambda}{m}, \label{a0}
\end{equation}
where $e$ is the electron charge, $E_0$ is the laser field strength in the focus, $\lambda$ is the wavelength, and $m$ is the electron (positron) mass. The discussion of physical properties, that we present, corresponds to $a_0\simeq 4000$  This value will be within the range of the next generation ultra intense pulsed lasers. For a plasma drop radius  $R = 3$ $\mu$m, $2R = 3\lambda$ the corresponding total plasma drop energy is  ${\cal O}(0.3)$\,kJ.

In the present context of plasma cooling we extended results of Ref.~\cite{Kuznetsova:2009bq}  to the lower density and lower temperature domain. The important  theoretical refinement discussed here for the first time, in the context of laser generated low density $e^-e^+$ plasma, is the consideration of the plasmon screening depending on plasma temperature and density. We also extend  our earlier  considerations to the nonrelativistic regime $T \le m$ as required in the study of the plasma expansion and freeze-out process.

Under the experimental conditions we consider here, all photons produced will escape from the small drop of low density plasma of electrons and positrons without much, if any, scattering. However, even  far from the chemical equilibrium density of the particle pair yield,  it is  possible for the produced electrons and positrons to  equilibrate thermally by means of 
M{\o}ller and Bhabha scattering, 
\begin{eqnarray}
e^{\pm}+e^{\pm} \leftrightarrow e^{\pm}+e^{\pm}\label{ee},\\ 
e^{\pm}+e^{\mp} \leftrightarrow e^{\pm}+e^{\mp}\label{ee1},
\end{eqnarray}
forming an electron-positron  plasma drop: wWhen the drop size $R$ exceeds the scattering length $L_{\rm ee}$,
\begin{equation}
R > L_{\rm ee}, \label{efreezout}
\end{equation}
multiple scattering processes can occur, allowing kinetic "`thermal"' equilibration. We therefore study positron annihilation loss processes assuming the Fermi-Boltzmann energy distribution of available particles. We solve kinetic population equations and evaluate the fraction of particles in plasma which can annihilate during the plasma life span.

There are two paths to positron annihilation, the direct in-flight pair annihilation,
\begin{equation}
e^{\pm}+e^{\mp} \rightarrow \gamma +\gamma
\end{equation}
and in-flight bound state positronium p$_s$ formation,
\begin{equation}
e^{\pm}+e^{\mp} \rightarrow \gamma + {\rm Pn}\qquad {\rm Pn}\to n\gamma,\  n=2,3
\end{equation}
which is followed ultimately by annihilation. The annihilation life span of positronium  for spin 0 is $\tau_{\rm P2}= 0.12$ ns, while for spin 1 it is $\tau_{\rm P3}=140$ ns. However, the positronium formation cross section only competes with the in-flight annihilation cross section for temperatures below  $T \approx 60$~eV~\cite{Gold1989}, and at that point, the expansion dilution will, in general, slow  these processes down considerably.  

In Sec.~\ref{kineq} we present cross sections for M{\o}ller and Bhabha scattering including in the plasmon screening effects. We compare the resulting pair annihilation cross section with positronium formation. In Sec. \ref{ld} we present numerical results for the M{\o}ller and Bhabha scattering mean free path and also annihilation relaxation time. We discuss conditions for plasma drop to be thermally equilibrated. In Sec. \ref{concl} we evaluate our results and present conclusions.

\section{$e^+, e^-$ Plasma reaction rates}\label{kineq}

\subsection{Scattering rates}
\subsubsection{Particle Density}
We consider the case of a small nonopaque expanding electron-positron plasma drop. The drop stays thermally equilibrated by scattering processes.   The electron (positron)  multiplicity $N_i$ ($i=e^+, e^-$) is  thus in thermal (momentum distribution) but not in chemical (yield distribution) equilibrium. 

It has been shown~\cite{Letessier:1993qa} that in order to maximize the entropy at fixed particle number  the appropriate maximum entropy distribution is the usual Fermi-Dirac  $f_{e,\bar e}$ distribution accompanied by a phase space occupancy parameter $\Upsilon$,
\begin{equation}\label{FDdis}
f_{e,\bar e} = \frac{1}{\Upsilon^{-1}e^{(u\cdot p\mp\nu)/T} +1}.
\end{equation}
$\Upsilon(t)$ describes the pair density and is, in general, a function of time, and it is the same for both particles and antiparticles. This is in contradistinction to the chemical potential $\nu$ which changes sign, $\nu_{\bar e}=-\nu_e$ comparing particles and antiparticles. The chemical potential $\nu$ regulates the abundance difference between particles and antiparticles and thus, in general, is only weakly dependent on time. A system with  $\Upsilon = 1$ for all particles  is  in chemical equilibrium,
and we refer to particle density with $\Upsilon = 1$  as a chemical equilibrium
density.

Note that the Lorentz-invariant exponents involve the scalar product of the  particle four-momentum $p^{\mu}_{i}$ with the local four-vector of velocity $u^{\mu}$, where  $u^{\mu}$ describes the local collective flow of matter, as expected for an unconfined plasma drop. The thermal properties $\nu,T,\Upsilon$ are defined in the local rest frame. In the absence of local matter  flow  the local  rest frame is the laboratory frame,
\begin{equation}\label{4v} 
u^{\mu}=\left(1,\vec{0}\right), \qquad p^{\mu}=\left(E, \vec{p}\right).
\end{equation}
We thus have 
\begin{equation}\label{FD}
f_{e,\bar e} = \frac{1}{\Upsilon^{-1}_{e,\bar e}e^{E/T} +1},\quad  \Upsilon_{e,\bar e}=\Upsilon e^{\pm\nu /T}
\end{equation}
The yields of particles  are
\begin{equation}
N_{e,\bar e} = n_{e,\bar e} V = g_{e,\bar e}V\int \frac{d^3p}{(2\pi)^3} f_{e,\bar e},\label{dens}
\end{equation}
where $V = 4\pi R^3/3$ is the volume and $g_{e,\bar e} = 2$ is the spin degeneracy. 
When the $e,\bar e$-pair yield is far below chemical equilibrium, that is,  $\Upsilon\ll 1$, the effects of quantum statistics are, in general, less significant and the Boltzmann limit is often equally precise, 
\begin{equation}\label{Bol}
f_{e,\bar e}  \to  \Upsilon_{e,\bar e} e^{-E/T} .
\end{equation}

\subsubsection{Plasmon mass and screening length}

\begin{figure}
\centering
\hspace*{0.4cm}\includegraphics[width=8.5cm,height=8cm]{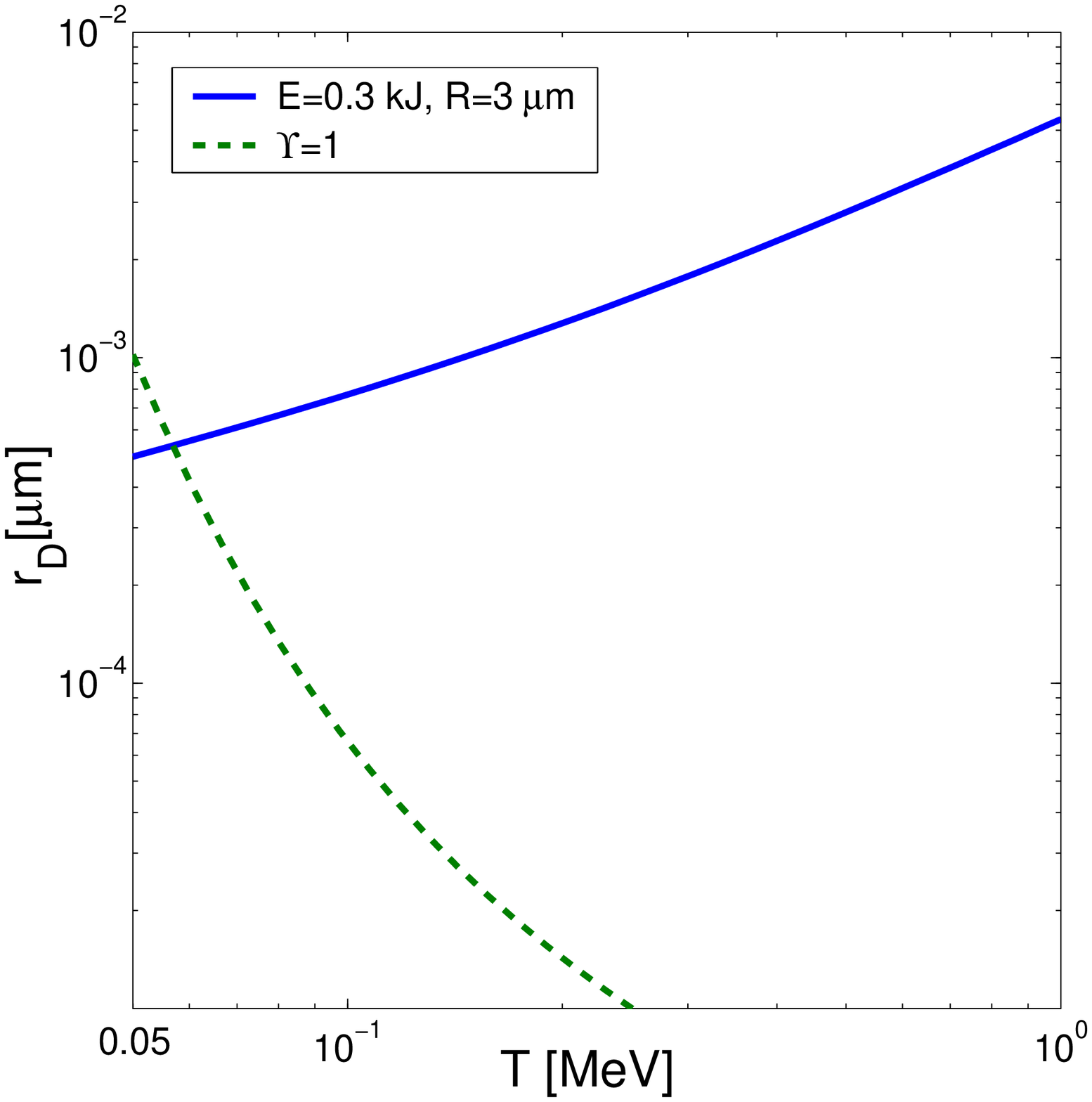}
\includegraphics[width=8cm,height=8cm]{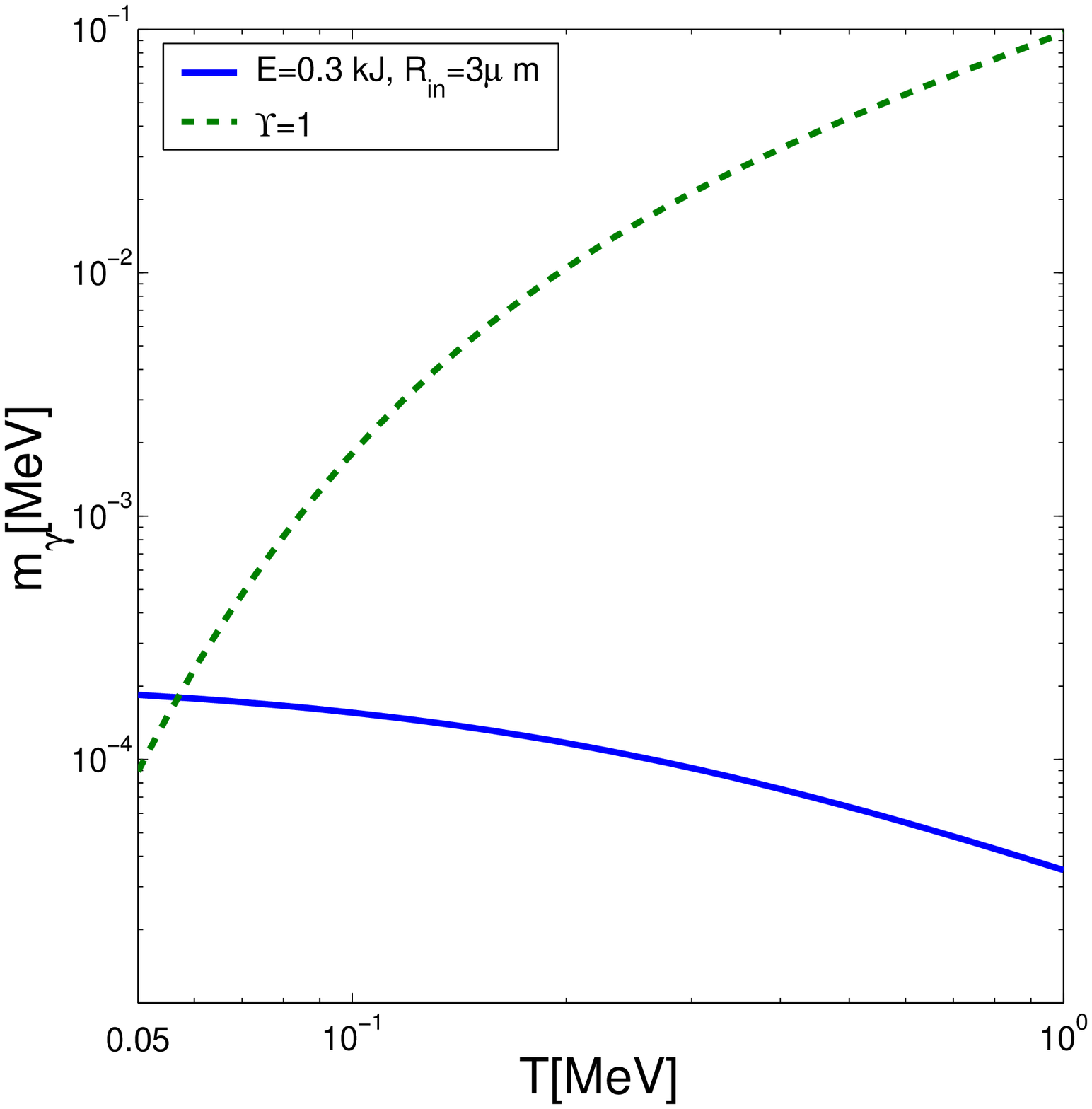}
\caption{\small{Upper panel: electron (positron) screening length  as a function of plasma temperature.  Lower panel: mass of the plasmon  as a function of $T$.}} \label{mthermal}
\end{figure}

To avoid Coulomb singularity in reaction matrix elements we introduce the plasmon mass, induced by the plasma screening effect, following the example of gluon dynamics in quark-gluon plasma~\cite{Biro:1990vj}. The plasmon mass is~\cite{Kislinger:1975uy}
\begin{equation}
m_{\gamma}^2 = \omega_{pl}^2 ={8\pi\alpha}\int\frac{f_{e^+}+f_{e^{-}}}{E_e}\left(1-\frac{p^2}{3E^2_e}\right)\frac{dp^3}{(2\pi)^3}. \label{mpl}
\end{equation} 
$\alpha = e^2/4\pi=1/137.036$ is the fine-structure constant.
For nonrelativistic temperatures $T<<m_e$, $m_{\gamma}$ goes to the classical plasma frequency, and a simple limit also emerges for relativistic temperatures with $\Upsilon = 1$,
\begin{equation}
m_{\gamma} \approx \left\{ 
\begin{array}{ll}
4\pi\sqrt{{2\alpha n_e}/{m_e}} & T<m_e, \\[0.3cm]
{\sqrt{4\pi\alpha}}T/{3} &T>m_e, \Upsilon = 1\\
\end{array}
\right.
\end{equation} 
The corresponding screening length, the Debye radius, is
\begin{equation}  
r_D = \frac{v_T}{\omega_{pl}}. \label{rD}
\end{equation}
and the mean thermal particle velocity $v_T$ is
\begin{equation}
v_T =  \large{\frac{\int \frac{p}{E} f  {d^3p} }
                   {\int f  {d^3p} }}
\end{equation}
since $f_{e^+}=f_{e^{-}}=f$.
We show in the Fig. \ref{mthermal} the electron (positron) screening length  and the mass of the plasmon as a function of $T$. 
The plasmon mass is increasing towards the small temperatures  and is asymptotically constant, similar to the behavior of the plasma density. 
The screening length is otherwise decreasing towards the small temperatures (inverse proportional to $m_{\gamma}$ and $v_T \propto \sqrt{T}$) in our range of temperature. 

\subsubsection{Boltzmann limit}

We are interested in experimental conditions under which the number of pairs produced is large compared to the residual electron density originating in matter. Furthermore we will deal with conditions ($\Upsilon_{\rm e} < 1$ or/and $T \leq m$ MeV) which allow us to use  the  Boltzmann approximation. Then, we have   
\begin{equation}\label{Bolrat}
\frac{n_e-n_{\bar e}}{n_e+n_{\bar e}}\to \sinh(\nu/T) \ll 1.
\end{equation}
In what follows we  will set $\nu =0$, and consider elsewhere the case for very low density degenerate plasma, where the chemical potential may become important.  We thus  have $\Upsilon_{e,\bar e}=\Upsilon$. In the relativistic Boltzmann  (classical) limit the plasma density and energy density are
\begin{eqnarray}
n_e &=& \frac{\Upsilon_{\rm e} g_eT^3}{2\pi^2} x^2K_2(x), \label{ne}\\
\epsilon &=& \Upsilon_{\rm e}\frac{3g_eT^4}{2\pi^2}\left(x^2K_2(x) + \frac{1}{3} x^3 K_1(x)\right),\label{enden}
\end{eqnarray}
where $K_i(x)$ is a Bessel function, $x=m/T$.

\subsubsection{Electron (positron) scattering rates}

In the evaluation of the matrix element we use Mandelstam  variables: s, u, and t. In the case of M{\o}ller scattering
\begin{equation}
s = (p_1+p_2)^2; \quad u = (p_3-p_2)^2; \quad t=(p_3-p_1)^2; \label{mv}
\end{equation}
and $s+u+t=m_1^2+m_2^2+m_3^2+m_4^2$. 

The M{\o}ller scattering matrix element is~\cite{Aksenov:2009dy, Halzen:1984mc, Kuznetsova:2009bq},
\begin{eqnarray}
&&\hspace{-0.6cm}
|M_{e^{\pm}e^{\pm}}|^{2}=2^{6}\pi^{2}\alpha^{2}\left\{
\frac{s^{2}+u^{2}+8m^{2}(t-m^{2})}{2(t-m^2_{\gamma})^{2}}  +\right.\notag\\
&&\hspace{-0.4cm}\left.
\frac{{s^{2}+t^{2}}+8m^{2}%
(u-m^{2})}{2(u-m_{\gamma}^2)^{2}} + \frac{\left( {s}-2m^{2}\right)\left({s}-6m^{2}\right)}
  {(t-m_{\gamma}^2)(u-m_{\gamma}^2)}\right\}. \label{M_fi_m}%
\end{eqnarray} 

In the case of Bhabha scattering we have
\begin{equation}
s = (p_3 - p_2)^2; \quad u = (p_1+p_2)^2; \quad t=(p_3-p_1)^2, \label{mvBh}
\end{equation}
see diagrams in~\cite{Kuznetsova:2009bq}.
The matrix element does not change in terms of variables $p_1, p_2, p_3$, when it is written in terms of variables $s, u, t$ we need to cross u and s in the M{\o}ller scattering matrix element [see Eq.\,(\ref{M_fi_m})], 
\begin{equation}
\left| M_{e^{\pm}e^{\mp}}(s,t,u)\right|^{2} 
= \left|M_{e^{\pm}e^{\pm}}(u,t,s)\right|^{2};\label{stcr}
\end{equation}
thus we find
\begin{eqnarray}
&&\hspace{-0.6cm}
 |M_{e^{\pm}e^{\mp}}|^{2}=2^{6}\pi^{2}\alpha^{2}
\left\{\frac{s^{2}+u^{2}+8m^{2}(t-m^{2})}{2(t-m^2_{\gamma})^{2}} +\right.\notag\\ &&\hspace{-0.4cm}\left.
\frac{u^{2}+t^{2}+8m^{2}%
(s-m^{2})}{2(s-m^2_{\gamma})^{2}}  +   \frac{\left({u}-2m^{2}\right)\left({u}-6m^{2}\right)}
 {(t-m^2_{\gamma})(s-m^2_{\gamma})} \right\}. 
\label{M_fi_b} 
\end{eqnarray}

For M{\o}ller and Bhabha scattering the cross section $\sigma_{ee}(s)$ can be obtained by averaging the matrix element over the $t$ variable:
\begin{equation}
\sigma_{ee}(s) = \frac{1}{16\pi(s-4m^2)^2}\int_{t_{\rm min}}^{t_{\rm max}} dt|M_{ee}|^{2}, 
\end{equation}
where $t_{\rm min} = -(s-4m^2)$, $t_{\rm max}=0$ in both cases~\cite{Kuznetsova:2009bq}.
Similar evaluations were done for heavy quarks production~\cite{Combridge:1978kx}.

\begin{widetext}
For M{\o}ller and Bhabha cross sections we obtain in plasma, keeping $m_\gamma$,
\begin{eqnarray}
\sigma_{e^{\pm}e^{\pm} \leftrightarrow  e^{\pm}e^{\pm}}(s)=
  \frac{1}{16\pi(s-4m^2)^2}\int_{-(s-4m^2)}^0 dt|M_{e^{\pm}e^{\pm}}|^{2} 
&=&
\frac{4\pi\alpha^2}{(s-4m^2)}\left(\frac{s^2+8m^2(m_{\gamma}^2-m^2)+(s+m_{\gamma}^2-4m^2)^2}{\left(s+m_{\gamma}^2-4m^2\right)m_{\gamma}^2}+1\right)\nonumber\\[0.4cm]
&&\hspace*{-5cm}+\frac{8\pi\alpha^2}{(s-4m^2)^2}\left(\frac{(s-2m^2)(s-6m^2)}{(s-4m^2+2m_{\gamma}^2)}+s+m_{\gamma}^2\right)\ln{\frac{m_{\gamma}^2}{s-4m^2+m_{\gamma}^2}}; \label{crmol}
\end{eqnarray}
\begin{eqnarray}
\sigma_{e^{\pm}e^{\mp} \leftrightarrow  e^{\pm}e^{\mp}}(s)=
  \frac{1}{16\pi(s-4m^2)^2}\int_{-(s-4m^2)}^0 dt|M_{e^{\pm}e^{\mp}}|^{2} 
&=&\frac{2\pi\alpha^2}{(s-4m^2)}\times\nonumber\\[0.4cm]
&&\hspace*{-6.5cm}
\left[\frac{s^2+8m^2(m_{\gamma}^2-m^2)+(s+m_{\gamma}^2-4m^2)^2}{\left(s+m_{\gamma}^2-4m^2\right)m_{\gamma}^2}+1 +\frac{8\left((s-4m^2)^2+m^2(s-m^2)\right)} {3(s-m_{\gamma}^2)^2}
         +\frac{3s+2m_{\gamma}^2+4m^2}{(s-m_{\gamma}^2)}\right.\nonumber\\[0.4cm]
&&\hspace*{-6.2cm}+ 2\left.\frac{(m_{\gamma}^2+s)^2-4m^4+(s^2-m_{\gamma}^4)}
{(s-m_{\gamma}^2)}\ln{\frac{m_{\gamma}^2}{s-4m^2+m_{\gamma}^2}}\right].
\label{crbh}
\end{eqnarray}
\end{widetext}

\subsection{$e+\bar e$ Annihilation}

\subsubsection{Master equation and annihilation time constant}
The master population equation  reads
\begin{equation}
\frac{1}{V}\frac{dN_{e,\bar e}}{dt} = -\Upsilon_{e}\Upsilon_{\bar e}  W_{\rm ann}.\label{ne0}
\end{equation}
We have made explicit the dependence of evolution of the  particle (pair) multiplicity in thin plasma on the prevailing density  showing the factor $\Upsilon_{e} \Upsilon_{\bar e}$.

A simplified form of the master equation (up to dilution by volume expansion, to be considered elsewhere) is easily obtained,
\begin{equation}
\frac{1}{\Upsilon_{e}}\frac{d\Upsilon_{e}}{dt}=-\frac{1}{\tau^e_{\rm ann}}
       \frac{\Upsilon_{\bar e}}{\Upsilon^{\rm in}_{\bar e}},
\end{equation}
introducing the  annihilation relaxation time $\tau^e_{\rm ann}$~\cite{Kuznetsova:2009bq}
\begin{equation}
\tau^e_{\rm ann}= \frac{dn_e/d\Upsilon_{\rm e}}{\Upsilon^{\rm in}_{e}W_{\rm ann}}.
\end{equation} 
and similarly for $\tau^{\bar e}_{\rm ann}$. In our case $\Upsilon_{e}\simeq \Upsilon_{\bar e}$ and we see that
\begin{equation}
\frac{\Upsilon^{\rm in}_{\bar e}}{\Upsilon_{\bar e}} 
    =\int_0^t\frac{dt'}{\tau^{\bar e}_{\rm ann}}(t')
\end{equation} 

We can write a similar master equation for the plasma drop energy loss,
\begin{equation}
\frac{1}{V}\frac{dE^{\rm tot}}{dt} = -\Upsilon_{e}\Upsilon_{\bar e}  W^E_{\rm ann},\label{E0}
\end{equation}
where $E^{tot}$ is the total energy of the plasma drop.
The relaxation time of energy loss is
\begin{equation}
\tau^E_{\rm ann}= \frac{d\epsilon/d\Upsilon_{\rm e}}{\Upsilon^{\rm in}_{e}W^E_{\rm ann}},
\end{equation} 
where $\epsilon$ is the plasma energy density and $\Upsilon^{\rm in}_{e}$ is the initial electron (positron) phase space occupancy.

\subsubsection{Annihilation rate in flight}
When electrons collide with positrons, they can annihilate. We consider here the dominant in flight annihilation process  into two photons. The invariant rate of annihilation per unit of volume and time $e+\bar e\to \gamma + \gamma$ is ($3+4\to 1+2$)
\begin{eqnarray}
&&W_{\rm ann}=\frac{g_{e}^2}{2(2\pi)^8}
\int\frac{d^{3}{p_{1}^{\gamma}}}{2E_{1}^{\gamma}} 
\int\frac{d^{3}{p_{2}^{\gamma}}}{2E_{2}^{\gamma}}
\int\frac{d^{3}{p_{3}^{e}}}{2E_{3}^{e}}
\int\frac{d^{3}{p_{4}^{\bar e}}}{2E_{4}^{\bar e}}
  \nonumber\\[0.2cm]
&&\times \delta^{4}\!\!\left(p_{1}^{\gamma} + p_{2}^{\gamma} 
                        - p_{3}^{e} - p_{4}^{\bar e} \right) 
\sum_{\rm spin}\left|\langle p_{1}^{\gamma}p_{2}^{\gamma}
\left| M_{\gamma\gamma \leftrightarrow e\bar e}\right|
p_{3}^{e}p_{4}^{\bar e}\rangle\right|^{2} \nonumber\\[0.1cm]
&&\times e^{u \cdot(p_{1}^{e} +p_{2}^{e} )/T}
f_{e}(p_{3}^{e})\Upsilon_{e}^{-1}f_{\bar e}(p_{4}^{\bar e})\Upsilon_{\bar e}^{-1}. \label{ggee}
\end{eqnarray}
Here $\langle p_{1}^{\gamma}p_{2}^{\gamma}\left| M_{\gamma\gamma \leftrightarrow e\bar e}\right| p_{3}^{e}p_{4}^{\bar e} \rangle$ is the annihilation quantum matrix element which we will consider to lowest order in $\alpha$, $g_e$ is electron-positron degeneracy, and factor $1/2$ is due to the indistinguishability of the final state photons.   We used this method  to describe the electron-positron pair  annihilation in~\cite{Kuznetsova:2009bq}, adapting it from work on   strangeness production in quark-gluon plasma~\cite{Rafelski:1982pu,Koch:1986hf,Matsui:1985eu,Koch:1986ud}. In the last line of 
Eq.(\ref{ggee}) we introduce  $\Upsilon_{e}^{-1} \Upsilon_{\bar e}^{-1}$ to compensate for the factor $\Upsilon_e Upsilon_{\bar e}$ seen in Eq.\,(\ref{ne0}).

The invariant rate Eq.\,(\ref{ggee})  relates to the 
electron-positron pair annihilation cross section~\cite{hadBook}; in the Boltzmann limit we have 
\begin{equation}
W_{\rm ann} = \frac{g^2 T}{32\pi^4} 
\int_{4m^2}^{\infty}ds \sqrt{s}(s-4m^2)\,\sigma_{ee \rightarrow \gamma\gamma}(s) K_1(\sqrt{s}/T). \label{ratebol}
\end{equation}
Here  the annihilation cross section is~\cite{Kuznetsova:2009bq}
\begin{eqnarray}
&&\hspace{-0.6cm}\sigma_{ee \rightarrow \gamma\gamma}(s) 
=\frac{2\pi\alpha^2(s^2+4m^2s-8m^4)}{s^2(s-4m^2)}\times\notag\\[0.3cm]
&&\hspace{-0.5cm}\left(
   \ln{\frac{\sqrt{s}+\sqrt{s-4m^2}}{\sqrt{s}-\sqrt{s-4m^2}}} 
   -\frac{\left(s+4m^2\right)\!\sqrt{s^2-4m^2s}}{(s^2+4m^2s-8m^4)}
\right).
\label{creegg}
\end{eqnarray}
 
\subsubsection{Energy loss}
Once in-flight $e+\bar e$ annihilation occurs, the produced photons escape the small plasma volume. An analogous expression to Eq.\,(\ref{ggee}) describes the energy loss rate due to pair annihilation, 
\begin{eqnarray}
&&W^E_{ann}=
\frac{g_{\gamma}^2}{2(2\pi)^8}\int\frac{d^{3}{p_{1}^{\gamma}}}{2E_{1}^{\gamma}} \int\frac{d^{3}{p_{2}^
{\gamma}}}{2E_{2}^{\gamma}}
\int\frac{d^{3}{p_{3}^{e}}}{2E_{3}^{e}}
\int\frac{d^{3}{p_{4}^{e}}}{2E_{4}^{e}}\times
  \nonumber\\[0.2cm]
&&\times \delta^{4}\left(p_{1}^{\gamma} + p_{2}^{\gamma} -
p_{3}^{e} - p_{4}^{e} \right) \sum_{\rm spin}\left|\langle
p_{1}^{\gamma}p_{2}^{\gamma}\left| M_{ee \rightarrow \gamma\gamma} \right|p_{3}^{e}p_{4}^{e}\rangle\right|^{2} \nonumber\\[0.2cm]
&&\times (E^{e}_3+E^{\bar e}_4) f_{e}(p_{3}^{e})f_{\bar e}(p_{4}^{e})\Upsilon_{e}^{-2}e^{u \cdot(p_{1}^{\gamma} + p_{2}^{\gamma} )/T},\label{ggeeE1}
\end{eqnarray}

We now obtain a relation analogous to Eq.\,(\ref{ratebol}). Consider the integral~\cite{hadBook} leading to  Eq.\,(\ref{ratebol}),
\begin{equation}
\int{d^4p e^{-\beta p\cdot u}\delta_{0}(p^2-s)} = \frac{2\pi}{\beta}\sqrt{s}K_1(\beta\sqrt{s}),
\end{equation}
where $u=(1,\vec 0)$ in the laboratory frame. Instead, we now need to use 
\begin{equation}
\int{d^4p \,p\cdot u\, e^{-\beta p\cdot u}\delta_{0}(p^2-s)}
=  -\frac{\partial}{\partial \beta}\frac{2\pi}{\beta}\sqrt{s}K_1(\beta\sqrt{s}).
\end{equation}
We use $d[K_1(x)/x]/dx=-K_2(x)/x$ to obtain
\begin{equation}
W^E_{ann}  =  \frac{g^2T}{32\pi^4} 
\int_{s_{th}}^{\infty}ds{s}(s-4m^2)\,\sigma_{ee\to \gamma\gamma}(s) K_2(\sqrt{s}/T).\label{ratebolE}
\end{equation}

\subsubsection{Positronium formation}

The cross section  for radiative positronium $(e\bar e)$ formation, $e^{-}+e^{+}\leftrightarrow \gamma+(e\bar e)$~\cite{Akhiezer1996} is
\begin{equation}
\sigma_{pos} = \frac{2^{12}\pi^2\omega}{3pm^2}\xi\left(\frac{\xi^2}{1+\xi^2}\right)^3\frac{e^{-4\xi \rm arccot \xi}}{1-e^{-2\pi\xi}}\left(1+\frac{\omega^2(1-\xi^2)}{5p^2}\right), \label{poscr}
\end{equation}
where $\xi = \alpha m/2p$ and the photon energy $\omega$ is defined by the conservation law
\begin{equation}
\omega+\frac{\omega^2}{4m} = p^2/m +\alpha^2 m/4.
\end{equation}
$p$ is the electron (positron) momentum in the center of mass reference frame, $p=\sqrt{s-4m^2}/2$. Equation (\ref{poscr}) is valid while $\xi \leq 1$. This condition is satisfied up to temperatures on the order of 10 eV.

We did not consider in detail the influence of plasma screening on positronium formation, a topic which invites further work in view of currently available results. It was found in~\cite{Jung2002} that the plasma screening and collective effects significantly reduce the radiative recombination cross section in non ideal plasma. The  screening effect for positronium formation should be similar to result for free electron radiative recombination with ions in nonideal classical plasmas. However, in positron - hydrogen plasma the Debye screening can result in a large increase of the positronium formation cross section at incident positron energy 20-100 eV~\cite{Sen2011}.

\section{Results for laser formed plasma}
\label{ld}

\subsection{Parameters for thermal plasma drop}

We assume here that the total energy $E$ of (colliding) laser pulses converts in the initial volume $V$  to the $e^+e^-$-plasma drop energy. The initial energy density $\epsilon=E/V$ is obtained from Eq.(\ref{a0}) and is characterized by $a_0$ and $\lambda$, 
\begin{equation}
\epsilon = \frac{1}{4\pi}E_0^2 = \frac{1}{4\pi}\left(\frac{a_0 m}{e\lambda}\right)^2.
\label{endenlas}
\end{equation}
The phase space occupancy of the plasma drop is 
\begin{equation}
\Upsilon_e = \frac{1}{4\pi \epsilon_0(T)}\left(\frac{a_0 m}{e\lambda}\right)^2, 
\label{upsil}
\end{equation}
where we introduced the chemical equilibrium energy density $\epsilon_0 = \epsilon|_{\Upsilon_e=1}$, Eq.(\ref{enden}). Then the total energy of plasma, $E$, is defined by the plasma drop radius $R$ for a given parameter $a_0$ and wavelength $\lambda$. The initial plasma size is expected to be close to the wavelength. We take the wavelength $R=3\lambda/2$ for all cases considered below.

\begin{figure}
\centering
\hspace*{0.35cm}\includegraphics[width=8.5cm,height=8cm]{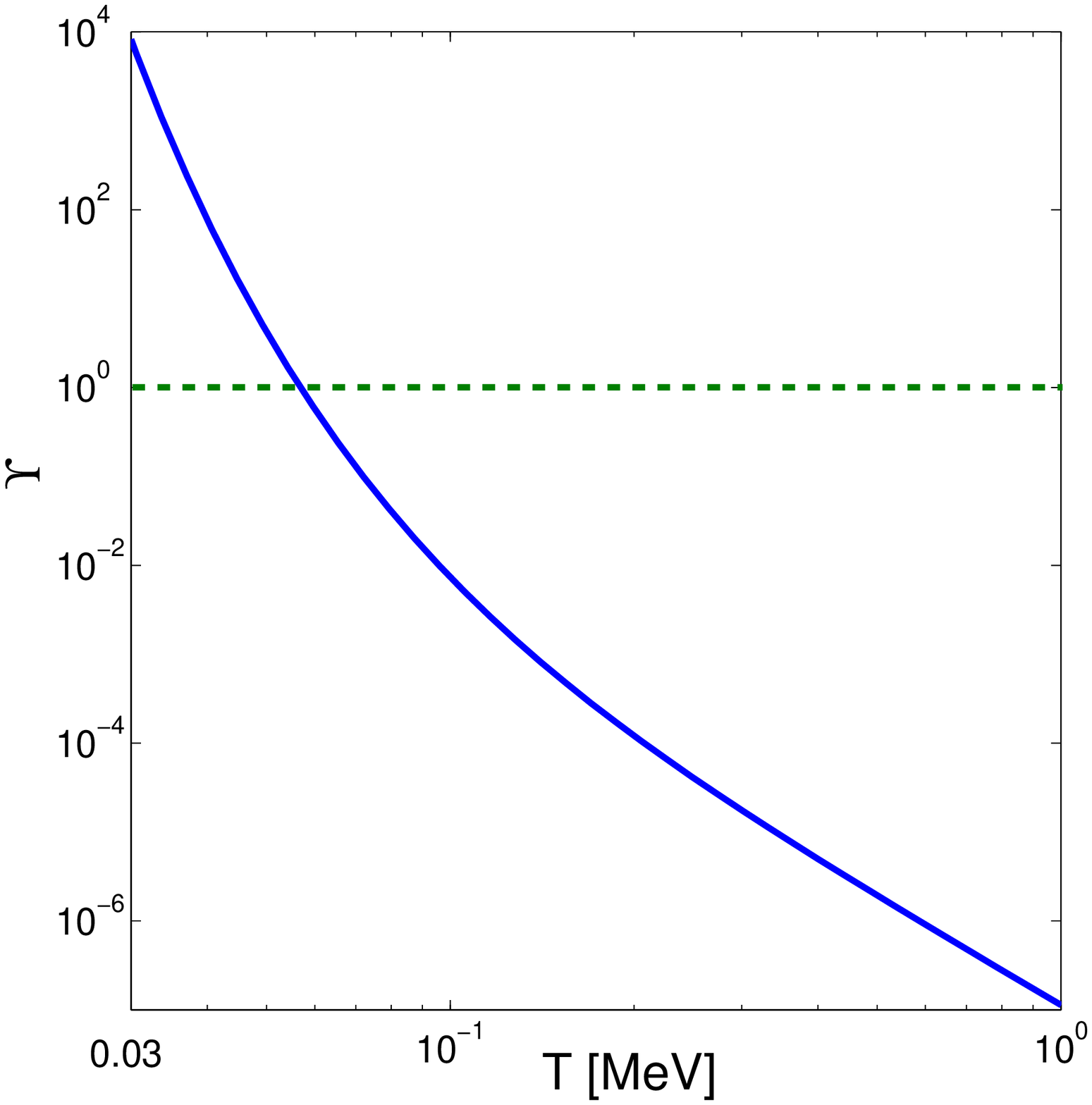}
\includegraphics[width=8cm,height=8cm]{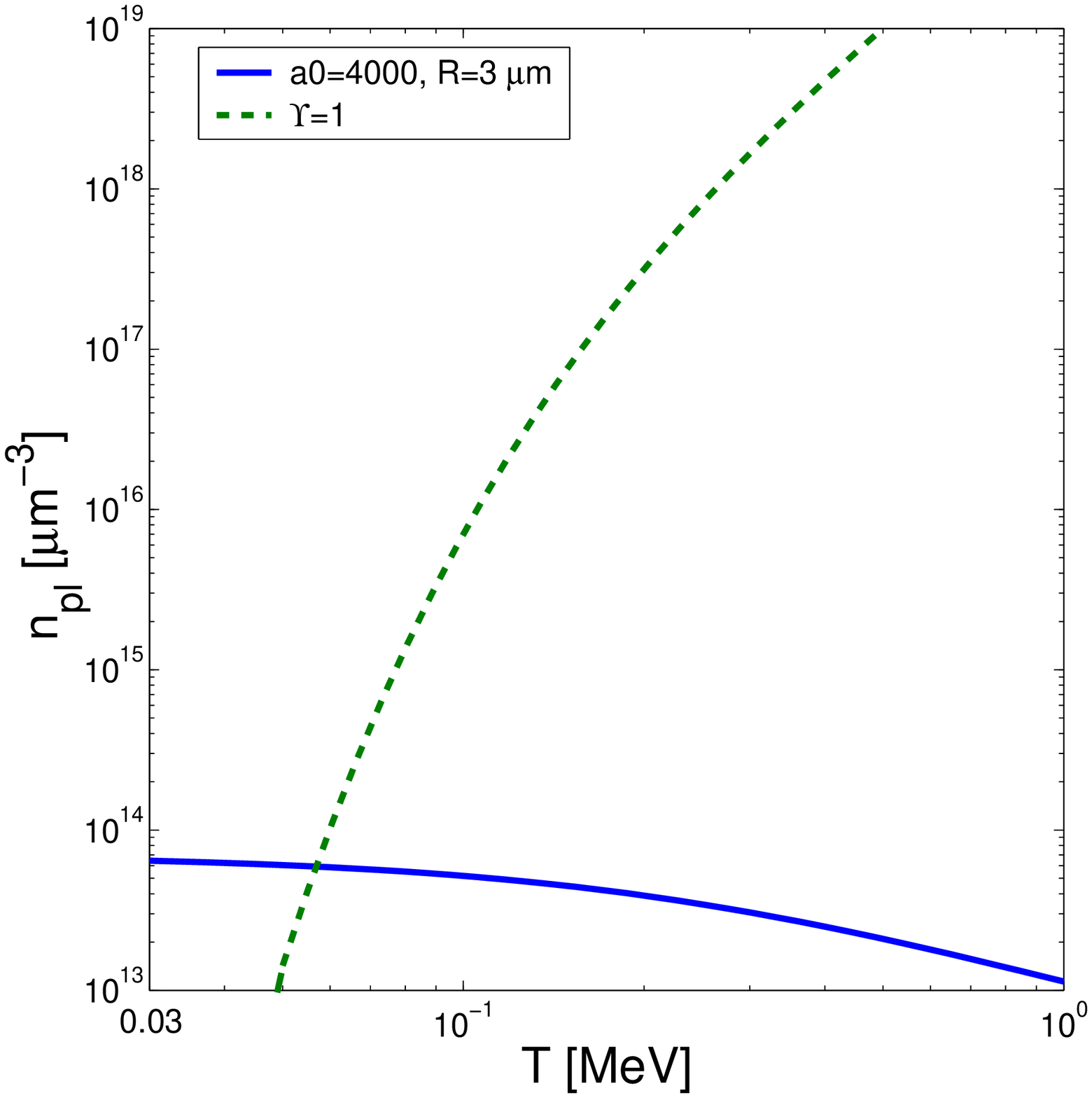}
\caption{\small{Upper panel: electron (positron) phase space occupancy $\Upsilon_e$ as a function of $T$ for  $a_0=4000$ and $R=3$ $\mu$m (solid blue line).  Lower panel: plasma density corresponding to the phase space occupancy on the upper panel (solid blue line) and equilibrium density $\Upsilon = 1$ (dashed green line) as a function of $T$.}} \label{neupsil} 
\end{figure}

In Fig.~\ref{neupsil} we show the phase space occupancy $\Upsilon_e$ from Eq.(\ref{upsil}) (upper panel) and the corresponding plasma density $n_{pl}$ (lower panel). The solid (blue) line shows the actual chemical nonequilibrium 
values. For comparison the chemical equilibrium results are shown by the dashed (green) line. We note that for $T>>0.06$ MeV the fully equilibrated yield is much greater than what we can make using a near future high intensity laser. However, the density of particles in plasma which we achieve is very high.

At $T<<m$, when plasma becomes nonrelativistic the  energy/particle $\to mc^2$ is a constant and does not depend much on the plasma temperature. Hence, the plasma particle density goes for $T\to 0$ to a constant for a given energy and plasma drop size,
\begin{equation}
n_{\rm pl}= n_e + n_{\bar e} = \frac{\epsilon}{mc^2}.
\end{equation}
and temperature cannot be determined considering a given available energy constraint.

\begin{figure}
\centering
\includegraphics[width=8cm,height=8cm]{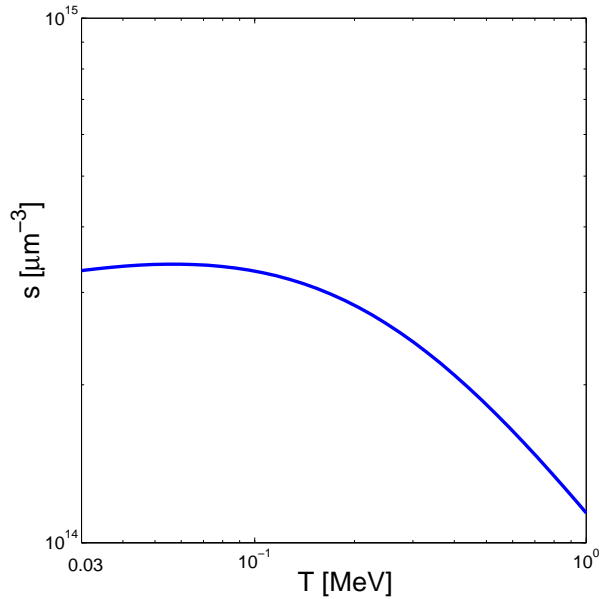}
\caption{\small{The entropy density of electron-positron plasma with $a_0=4000$ and $R = 3$ $\mu$m as a function of temperature.}} \label{entdens}
\end{figure}
In a system  where   particle (pairs) can be produced but energy is fixed the entropy density reaches maximum at $\Upsilon = 1$. We show the entropy density of electron - positron plasma,
\begin{equation}
s = \int \frac{d^3p}{2\pi^3}\left((f_e-1)\ln(1-f_e)-f_{e}\ln(f_e)\right),
\end{equation}
at $E=0.3$ kJ and $R = 3$ $\mu$m as a function of temperature in Fig.~\ref{entdens}. As expected, the maximum of the entropy density is at the temperature, $T=0.06$ MeV, where phase space occupancy of electron and positron $\Upsilon_e = 1$. However, the maximum is very flat. Note that there is much less entropy density  when the system is formed at relatively high temperature. This is because there are fewer particle pairs and, for a relativistic gas, the entropy per particle is near $S/N\simeq 4$. For far off equilibrium low density systems the expansion of the volume is thus accompanied by reactions that tend to chemically equilibrate the system and move it towards   chemical equilibrium.

\subsection{Electron and positron scattering}

The formation of electron-positron plasma is further subject to the opacity condition Eq.(\ref{efreezout}). To check if this condition is satisfied  we extend our earlier considerations~\cite{Kuznetsova:2009bq},  now introducing plasmon mass, Eq.(\ref{mpl}), in a domain of  mild relativistic and nonrelativistic temperatures.

The electron (positron) mean free path follows from
\begin{equation}
L_{\rm ee} = \frac{n_e}{W_{\rm ee}},\label{sclee1}
\end{equation}
where for the scattering rate $W_{\rm ee}$ we use an equation similar to Eq.(\ref{ratebol}) (since the final state does not have two identical bosons, the normalization factor is different):
\begin{equation}
W_{\rm ee} = \frac{g^2 T}{32\pi^4} 
\int_{4m^2}^{\infty}ds \sqrt{s}(s-4m^2)\,\sigma_{\rm ee}(s) K_1(\sqrt{s}/T),\label{ratebol1}
\end{equation}
and
\begin{equation}
\sigma_{\rm ee} = \sigma_{e^+e^+ \leftrightarrow e^{-}e^{-}} + \sigma_{e^-e^+ \leftrightarrow e^+e^-}.
\end{equation}

\begin{figure}
\centering
\includegraphics[width=8cm,height=8cm, height=8.5cm]{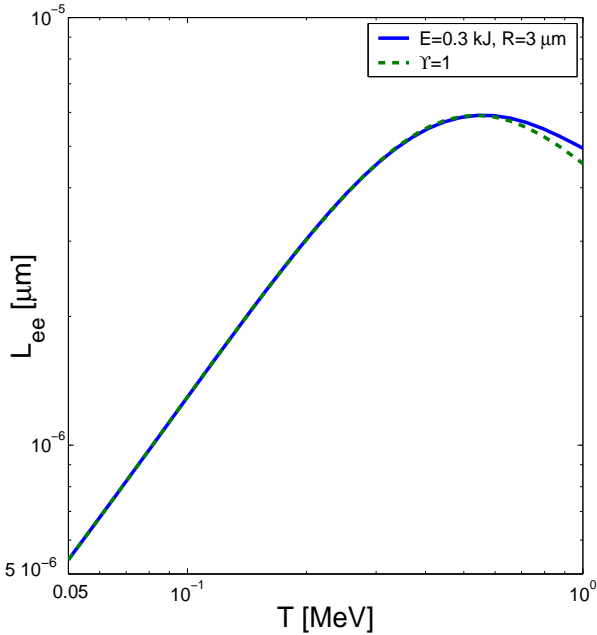}
\caption{\small{Electron (positron) scattering length at a given plasma radius and energy as a function of  $T$.}} \label{lsc}
\end{figure}

In Fig.~\ref{lsc} we show the electron (positron) scattering length $L_{\rm ee}$, Eq.(\ref{sclee1}), at a given plasma radius $R = 3$ $\mu$m and energy 0.3 kJ ($a_0 = 4000$) as a function of plasma temperature  $T$.  $\Upsilon$ varies for every value of $T$, as we see in Fig. \ref{neupsil}. Since $\Upsilon_{\rm e} << 1$  the scattering length can be evaluated in the Boltzmann limit in practically the entire temperature range of interest, including $T>m$.  We also show  (dashed green line), for comparison, the case $\Upsilon_{\rm e} = 1$, which means that we allow the density to go up significantly and the small difference we see in figure~\ref{lsc} for high $T$ is due to quantum gas properties.  

At relativistic temperatures  $T \simeq  1$ MeV our present result is in agreement with scattering   rates evaluated with plasmon mass taken in the limit of ultrarelativistic temperatures in~\cite{Kuznetsova:2009bq} with an accuracy of few percent. 

For the constant plasma drop energy  scattering length $L_{\rm ee}$ has a maximum at $T \approx m$ . In the whole temperature range the plasmon mass is small and the first term in Eq.\,(\ref{crmol}) and Eq.\,(\ref{crbh}) is dominant, resulting in the cross section for electron or positron scattering,
\begin{equation}
\sigma_{\rm ee} \propto m_{\gamma}^{-2} \propto n_{\rm e}^{-1}. \label{crsn}
\end{equation} 
In the range where condition (\ref{crsn}) is valid, the electron (positron) mean free path does not depend on density or $\Upsilon_e$. When the mean free path is increasing with decreasing density, this is compensated by a larger cross section because of a smaller plasma screening effect or smaller $m_{\gamma}$. For the entire $T$ range, the scattering length scale is a tiny fraction of the plasma size. 

In the temperature range  $T<m$ the contribution of $4p^2 = s-4m^2$ is much smaller than $m^2$ and much larger than $m_{\gamma}^2$; the approximate cross sections for M{\o}ller and Bhabha scatterings, Eq.\,(\ref{crmol}) and Eq.\,(\ref{crbh}), are
\begin{equation}
\sigma_{e^{\pm}e^{\pm} \leftrightarrow  e^{\pm}e^{\pm}}(s) = 2\sigma_{e^{\pm}e^{\mp} \leftrightarrow  e^{\pm}e^{\pm}}(s) = 
\frac{64\pi\alpha^2}{(s-4m^2)^2}\frac{m^4}{m_{\gamma}^2}.\label{lTcrs}
\end{equation}
 
One can also consider a Rutherford-type differential cross section for M{\o}ller scattering ~\cite{Halzen:1984mc} 
\begin{equation}
\frac{d\sigma}{d\cos\theta} = \frac{\pi\alpha^2 m^2}{4p^4} {\rm cosec}^4 \theta/2. \label{rcr}
\end{equation}
We checked by integrating Eq.(\ref{rcr}) numerically that Eq.(\ref{lTcrs}) corresponds to the total cross section from the integrated Eq.(\ref{rcr}) with a cutoff angle $\theta_{\rm min} = m_{\gamma}/m$.

We found from the results presented in Fig.~\ref{lsc} that condition \ref{efreezout} is satisfied for whole temperature range considered. We conclude that the electron-positron plasma drop can stay thermally equilibrated at relatively low densities when $\Upsilon << 1$ and/or the temperature $T << m$:  the electron-positron mean free path decreases when the temperature decreases below the electron mass because of the factor $s-4m^2 =4 p^2$ in the denominator of cross section Eq.\,(\ref{lTcrs}). At a temperature higher than $m$ the other terms begin to contribute to the cross sections, Eq.\,(\ref{crmol}) and Eq.\,(\ref{crbh}). The electron-positron mean free path decreases again.  

The  cross section,~Eq.\,(\ref{lTcrs}), is valid in the temperature range
\begin{equation}
 T_{cr} = \frac{2\pi\alpha n_e}{m^2} < T < m, \label{Tcr}
\end{equation}
The reader should keep in mind that the present considerations do not
automatically apply to the case of a degenerate electron-positron
gas (high density or/and low temperature), where we should extend
the investigation of collective plasmon dynamics in order to obtain
a valid estimate of the electron-positron scattering cross section.

\subsection{Annihilation}
\subsubsection{Plasmons}
While screening and plasma oscillations impact the scattering processes, this is not the case for our domain in regard to the annihilation process.   There are several processes to consider:
\begin{enumerate} 
\item
The electron (positron) thermal mass correction, which is on the order of magnitude of $m_{\gamma}$. However,  $m_{\gamma} << m_e$ and this correction is small.
\item 
${\rm Plasmon} \leftrightarrow e^+e^-$,  if the reaction threshold is exceeded, $m_{\gamma} > 2m_e$~\cite{Biro:1990vj}.  This can only happen at ultrarelativistic temperatures. In the case considered here with constant plasma energy,  $m_{\gamma}^2 \propto T^{-1}$ (see Fig. \ref{mthermal}),  the threshold condition cannot be satisfied.
\item
The hard photons from annihilation $(k \approx m)$ are rescattering on plasmons. The condition where screening has noticeable effect on the photon propagation is~\cite{Glenzer2010}
\begin{equation}
kr_D \leq 1,
\end{equation}
where $k$ is the photon wave number and $r_D$ is the Debye radius Eq.(\ref{rD}). 
This condition is equivalent to the condition $T<T_{cr}$, Eq.(\ref{Tcr}). We do not consider here such low temperature plasma.
\end{enumerate}

\subsubsection{Annihilation life span}

We determine, using the perturbative QED reaction rate, the annihilation rate of plasma under the conditions considered in the previous subsections.  We assume that the plasma drop formation life span is on the order of magnitude of the laser pulse duration, 10 fs, and this is the stage at which the density of pairs and thus annihilation should have the largest rate; however, this is not the case, since, as $T$ increases, the pair density drops, given the constant initial total  energy, and thus the annihilation relaxation time increases. 

\begin{figure}
\centering
\vspace*{-0.5cm}\includegraphics[width=8cm,height=8cm, height=8.5cm]{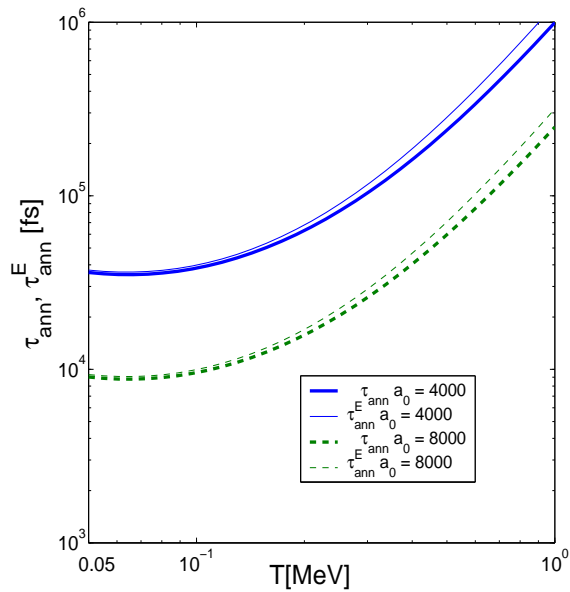}
\caption{\small{ Time constant for particle annihilation (thick lines) and energy loss (thin lines)  at $a_0 = 4000, E=0.3$\,kJ (solid blue lines) and $a_0=8000, E=1.2$\,kJ (dashed green lines) as a function of plasma drop temperature.}} \label{eannih}
\end{figure}

In Fig.  \ref{eannih} we show relaxation times $\tau$ for particle number annihilation  $\tau_{\rm ann}$  (thick lines) and energy loss  $\tau_{\rm ann}^{\rm E}$ (thin lines)  for plasma at $a_0 = 4000$, $E = 0.3$ kJ (solid blue lines) and $a_0 = 8000$, $E= 1.2$ kJ (dashed green lines) as a function of temperature. The values of $\tau$ are indeed largest for initial highest temperatures and there is a shallow minimum at $T \approx 0.065$ MeV.   At $T < 0.065$ MeV the pair density is approximately constant but particle temperature decrease results to increase of annihilation relaxation time. The fastest annihilation occurs here   because we have at this low temperature the highest mobility of particles at high density.

We recognize that the fraction of annihilations is very small initially, we  obtain from Eq.(\ref{ne0})
\begin{equation}
N_{\rm ann}/N_{0} \approx \Upsilon_{\rm e}^2 W_{\rm ann} \frac{t}{n_0} \approx \frac{t}{\tau_{\rm ann}}.
\end{equation}
Another way to look at the conditions fo annihilation is to note that the relaxation time is inversely proportional to $\Upsilon_{\rm e}$. Then from Eq.(\ref{endenlas}) we have 
\begin{equation}
\tau_{\rm ann} \propto \frac{\lambda^2}{a_0^2},
\end{equation}
which explains the dependence on $a_0$ (see Fig. \ref{eannih}).

We see in Fig. \ref{eannih} that the energy loss relaxation time $\tau^E_{ann}$ becomes very close to $\tau_{ann}$ for $T<m$, since the energy of the plasma drop changes mostly because of the pair mass disappearance and the resulting decrease in plasma mass . At   $T>2m$, the energy loss relaxation time is, as expected,  above the annihilation relaxation time. This happens since there is a preference for slower particles to annihilate, and thus on average, in the thermal bath few particles of higher energy lost and annihilation leads to a slight increase of the ambient plasma temperature.
 
Our result seen in Fig. \ref{eannih} implies that the annihilation process, even at the highest initial density, is relatively slow compared to other dynamical effects controlling the plasma drop:  the plasma drop must live   $t >> \tau_{ann}$ to have most positrons in the plasma  annihilated. This time is much longer than the pulse duration, 10 fs; indeed, it is on the scale of nano seconds. There is, furthermore, the kinetic  expansion leading to further dilution of the plasma -- when the plasma drop expands with time, the density decreases and the annihilation relaxation becomes even longer. Most, if practically not all of the $3\, \times 10^{-4}-10^{-5}$ annihilation events originate in the   densest plasma stage during laser pulse, and a reliable prediction of the total annihilation yield requires detailed control of the kinetic processes in the initial state of the plasma as well as a precise understanding of the plasma drop expansion dynamics, which further reduces the annihilation rate, ultimately leading to a cloud of streaming electrons and positrons. 

\subsubsection{In-flight annihilation compared to positronium formation}

\begin{figure}
\centering
\includegraphics[width=8cm,height=8cm]{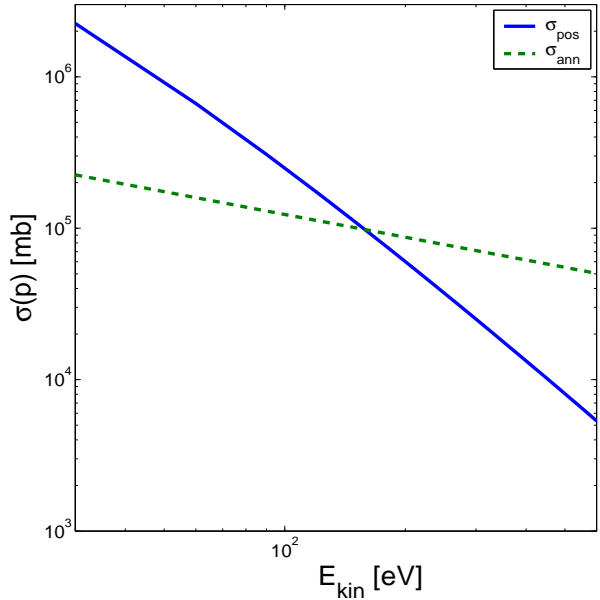}
\caption{\small{Radiative positronium formation (solid line) and direct annihilation cross sections (dashed line) as a functions of electron (positron) kinetic energy in the two particle center of momentum frame.}} \label{posit}
\end{figure}

In Fig.~\ref{posit} we compare the nonrelativistic limit of the annihilation in-flight cross section (dashed line)  to the cross section  for radiative positronium $(e\bar e)$ formation (solid blue line)  as a function of electron(positron) kinetic energy in the center of mass frame $E_{\rm kin} = {(s-4m^2)}/8m$.  The cross sections intersect at $E_{\rm kin} \approx 150$ eV. This corresponds to the crossover temperature obtained in~\cite{Gold1989}, $T_e\simeq 60$ eV. Thus the direct annihilation dominates down to this low temperature, and our prior results apply for $T>T_e$. For $ T<T_e$ we have significant positronium formation only if we reach this condition without much of expansion, which is not part of our present study.

\section{Conclusions}\label{concl}

The key result of this study is that high intensity QED cascading leads to an electron-positron  drop which does not annihilate but   thermally equilibrates. In this plasma drop electron - positron pairs are thermalized by M{\o}ller  [Eq.(\ref{ee})] and Bhabha [Eq.(\ref{ee1})] scattering,  and they annihilate very slowly; see Fig. \ref{eannih}. 

We found that in the Boltzmann limit the electron and positron scattering length nearly does not depend on plasma density in considered temperature range due to collective plasmon effects.  The cross section decrease at lower density is compensated by plasmon charge screening in the less dense plasma. As a result electron-positron plasma can be thermally equilibrated at the density and temperature range considered, far below the chemical equilibrium of the pair yield, $\Upsilon =1$. The plasma drop size allows very many scattering processes; we did not find any restriction on the minimum plasma drop energy and/or maximum drop size by considering the opaqueness condition Eq.(\ref{efreezout}) for electron (positron) scattering. 

We calculated, as an example, the annihilation relaxation time for an internal plasma drop energy of 0.3 -- 1.2  kJ and radius 3 $\mu$m. Because of the relatively low density the annihilation relaxation time is much longer than the pulse duration, which is $\approx 10$ fs. We found that in-flight annihilation is fastest at $T=0.065$ MeV, yet still relatively slow. The radiative positronium production   process   exceeds the in-flight annihilation at a much lower  temperature, 60 eV, leading perhaps to the formation of  positronium in the late stages of the drop. If such a low temperature is reached without drastic expansion dilution, very many positroniums can be  formed,  and  positronium formation prolongs the life span of positrons, though the nature  of the plasma drop is now different. 

The experimental conditions will determine at what temperature and, more importantly for the following  argument, rapidity, relative to the laboratory frame of reference, the electron-positron drop will be formed~\cite{Labun:2011xt,Labun:2011pj}. Multipulse arrangements can be easily obtained, resulting in the plasma drop being formed at high rapidity. The greater the rapidity, the greater the effect of the time dilation that prolongs the life span of the plasma drop, as seen in the laboratory. We recognized, in this work, the relative stability against annihilation evaluated in the intrinsic rest frame of the drop. Therefore, it appears possible to create, using high density lasers, a quasistable matter-antimatter plasma drop capable of traveling macroscopic distances before dissipating into a low density cloud of particles.

\begin{acknowledgments}   
This work was supported in part by the
U.S. Department of Energy Grant No. DE-FG02-04ER41318
\end{acknowledgments}

\vspace*{-0.3cm}

\end{document}